\documentclass[a4paper,twoside,twocolumn,english,british,journal]{IEEEtran}
\usepackage[T1]{fontenc}
\usepackage[latin9]{inputenc}
\usepackage{babel}
\usepackage{units}
\usepackage{mathtools}
\usepackage{amsmath}
\usepackage{amssymb}
\usepackage{graphicx}
\usepackage{esint}
\usepackage[unicode=true,
 bookmarks=true,bookmarksnumbered=true,bookmarksopen=true,bookmarksopenlevel=1,
 breaklinks=false,pdfborder={0 0 0},pdfborderstyle={},backref=false,colorlinks=false]
 {hyperref}
\hypersetup{pdftitle={Model-Free Closed-Loop Stability Analysis: A Linear Functional Approach},
 pdfauthor={Adam Cooman, Fabien Seyfert, Martine Olivi, Sylvain Chevillard and Laurent Baratchart},
 pdfpagelayout=OneColumn,pdfnewwindow=true,draft}

\makeatletter

\pdfpageheight\paperheight
\pdfpagewidth\paperwidth

 \let\oldforeign@language\foreign@language
 \DeclareRobustCommand{\foreign@language}[1]{%
   \lowercase{\oldforeign@language{#1}}}

\usepackage{color}
\usepackage{babel}
\usepackage{units}
\usepackage{bm}
\usepackage{dblfloatfix}
\usepackage{siunitx}

\usepackage{pgfplots}
\pgfplotsset{compat=newest} 
\pgfplotsset{plot coordinates/math parser=false,
every axis/.append style={font=\footnotesize},
every axis plot/.append style={join=round},
every x tick label/.append style={font=\footnotesize},
every y tick label/.append style={font=\footnotesize}
} 
 
\newcommand{\tikzify}[1]{\protect\tikz[baseline=-0.55ex]{\protect\draw #1}}

\newcommand{\marker}[2]{\tikzify{plot[mark=#2,mark size=2.,mark options={color=#1}](0,0)}}

\newlength\figureheight 
\newlength\figurewidth 
\usetikzlibrary{plotmarks}
\usepackage{grffile}
\usepackage{cite}

\usepackage[toc,acronym,section=section,nonumberlist,xindy]{glossaries}

\renewcommand{\newacronym}[4][]{\newglossaryentry{#2}{type=\acronymtype,
name={#3},
short={#3},
long={#4},
shortplural={#3\glspluralsuffix},
longplural={#4\glspluralsuffix},
description={#4},
text={#3},
first={#4~(#3)},
plural={#3\glspluralsuffix},
firstplural={#4\glspluralsuffix~(#3\glspluralsuffix)},
#1}}
\newacronym{BLA}{BLA}{Best Linear Approximation}
\newacronym{MIMO}{MIMO}{Multiple-Input Multiple-Output}
\newacronym{SISO}{SISO}{Single-Input Single-Output}
\newacronym{SIMO}{SIMO}{Single-Input Multiple-Output}
\newacronym{PISPO}{PISPO}{Period-In Same Period-Out}
\newacronym{FRF}{FRF}{Frequency Response Function}
\newacronym{DFT}{DFT}{Discrete Fourier Transform}
\newacronym{FFT}{FFT}{Fast Fourier Transform}
\newacronym{HB}{HB}{Harmonic Balance}
\newacronym{ADS}{ADS}{Advanced Design System}
\newacronym{LSSS}{LSSS}{Large-Signal Small-Signal}
\newacronym{HTF}{HTF}{Harmonic Transfer Function}
\newacronym{RMS}{rms}{root mean square}
\newacronym{CAD}{CAD}{Computer-Aided Design}
\newacronym{OPAMP}{op-amp}{Operational Amplifier}
\newacronym{NDF}{NDF}{Normalised Determinant Function}
\newacronym{LTI}{LTI}{Linear Time-Invariant}
\newacronym{LTP}{LTP}{Linear Time-Periodic}
\newacronym{LF}{LF}{Low-Frequency}
\newacronym{HF}{HF}{High-Frequency}
\newacronym{RF}{RF}{Radio Frequency}
\newacronym{RR}{RR}{Return Ratio}
\newacronym{LNA}{LNA}{Low Noise Amplifier}
\newacronym{SSB}{SSB}{Single-Sideband}
\newacronym{VUB}{VUB}{Vrije Universiteit Brussel}
\newacronym{PA}{PA}{Power Amplifier}
\makeglossaries

\def\figurename{ADAM.}

\usepackage{flushend}

\makeatother

\begin{document}
\global\long\def\jw{\!\left(j\omega\right)}
\global\long\def\imp{Z_{mn}\jw}
\global\long\def\s{\left(s\right)}
\global\long\def\fmax{f_{\mathrm{max}}}
\global\long\def\fmin{f_{\mathrm{min}}}
\global\long\def\Zstable{Z_{\mathrm{stable}}\jw}
\global\long\def\Zunstable{Z_{\mathrm{unstable}}\jw}
\global\long\def\filtered{_{\mathrm{f}}}
\global\long\def\Zfiltered{Z\filtered\jw}
\global\long\def\filter{H\jw}
\global\long\def\jt{\left(\theta\right)}
\global\long\def\disc{^{\mathrm{disc}}}
\global\long\def\Zdisc{Z\filtered\disc\jt}
\global\long\def\Rstab{R_{\mathrm{stab}}}
\global\long\def\ex{\mathrm{ex}}

\def\figurename{Fig.}

\markboth{Longer version of paper published in IEEE Tran. on Microwave Theory
and Techniques. DOI: \href{http://10.1109/TMTT.2017.2749222}{10.1109/TMTT.2017.2749222}}{Cooman, Seyfert \MakeLowercase{\emph{et al}}\emph{.}: Model-Free
Closed-Loop Stability Analysis: A Linear Functional Approach}

\title{Model-Free Closed-Loop Stability Analysis:\\
A Linear Functional Approach}

\author{Adam~Cooman,~\IEEEmembership{Member,~IEEE}, Fabien~Seyfert, Martine~Olivi,
Sylvain~Chevillard and Laurent~Baratchart\thanks{Adam~Cooman, Fabien~Seyfert, Martine~Olivi, Sylvain~Chevillard
and Laurent~Baratchart are with APICS at INRIA, Sophia Antipolis.
e-mail: adam.cooman@inria.fr}}
\maketitle
\begin{abstract}
Performing a stability analysis during the design of any electronic
circuit is critical to guarantee its correct operation. A closed-loop
stability analysis can be performed by analysing the impedance presented
by the circuit at a well-chosen node without internal access to the
simulator. If any of the poles of this impedance lie in the complex
right half-plane, the circuit is unstable. The classic way to detect
unstable poles is to fit a rational model on the impedance. 

In this paper, a projection-based method is proposed which splits
the impedance into a stable and an unstable part by projecting on
an orthogonal basis of stable and unstable functions. When the unstable
part lies significantly above the interpolation error of the method,
the circuit is considered unstable. Working with a projection provides
one, at small cost, with a first appraisal of the unstable part of
the system. 

Both small-signal and large-signal stability analysis can be performed
with this projection-based method. In the small-signal case, a low-order
rational approximation can be fitted on the unstable part to find
the location of the unstable poles.
\end{abstract}

\IEEEpeerreviewmaketitle{}

Frequency domain simulation methods, like \gls{HB} or a DC analysis,
impose a structure on the obtained solution of the circuit \cite{Suarez2015}:
The DC analysis only allows for a fixed solution, while \gls{HB}
imposes a frequency grid. Any circuit solution that requires more
than the imposed frequencies, e.g. an extra oscillation not on the
imposed grid, cannot be represented in the constrained frequency grids
of DC and \gls{HB}. The simulator will still find a valid solution,
but the obtained orbit will be locally unstable: it cannot recover
from small perturbations and will be physically unobservable in the
circuit \cite{Suarez2002}. It is therefore necessary to perform a
local stability analysis on each of the circuit solutions obtained
with a DC and \gls{HB} analysis \cite{Suarez2015}.

Over the years, several methods have been developed to determine the
local stability of a circuit solution. Some techniques, like the analysis
of the characteristic system \cite{Suarez2002}, require access to
the simulator. Open-loop techniques, like the analysis of the normalised
determinant \cite{Suarez2015}, require access to the intrinsic device
models. These classic techniques are therefore hard to implement in
commercial simulators.

Closed-loop stability analysis methods can easily be applied as a
post-processing step without any internal knowledge of the circuit
and can be used in commercial simulators. This is the reason why they
have attracted a large interest lately~\cite{Suarez2015,Jugo2001,Collantes2004}.
A closed-loop local stability analysis performs linearisation of the
circuit around the orbit to check the stability thereof: if the linearised
circuit has at least one pole in the complex right half-plane, the
orbit is unstable. It is moreover assumed that, conversely, the absence
of unstable pole implies stability, although no published proof of
this fact seems available yet. The question is more subtle than it
looks: there exist delay systems which are unstable and still their
transfer-function has no unstable pole \cite{Partington}, moreover
has an example of an ideal circuit with this property. Nevertheless,
it is claimed in \cite{Baratchart2014} that a circuit whose elements
are passive at arbitrary high frequencies must indeed have some unstable
pole if it is unstable.

The poles of the linearisation around the circuit orbit cannot be
obtained directly. Instead a FRF of the linearised circuit is obtained
with small-signal simulations on a discrete set of frequencies. The
closed-loop stability analysis then aims at determining whether the
underlying FRF has a pole in the complex right half-plane. In a pole-zero
stability analysis, a rational approximation is fitted on the FRFs.
If the rational approximation contains poles in the complex right
half-plane, the solution is declared unstable. 

Note that the FRF of circuits with distributed elements, like transmission
lines, is not rational. Therefore it must be argued that the poles
of the computed rational approximant convey information on the poles
of the true FRF. This is a delicate issue and a particular instance
of a recurring question in approximation theory, namely: what do the
singularities of an approximant tell us about the singularities of
the approximated function? We observe that no such information can
be drawn from the mere quality of approximation in a range of frequencies,
since a famous theorem by Runge entails that a continuous function
on a segment can be approximated arbitrary well by a proper rational
function with prescribed pole location \cite{Rudin}. Thus, for singularity
detection, the choice of the approximation algorithm (and not just
the fit of the approximant) does matter. 

For instance, methods based on linear interpolation, like Padé or
multipoint Padé approximation, are famous for generating spurious
poles that wander about the domain of analyticity of the approximated
function. This phenomenon was intensively studied for meromorphic
and branching functions \cite{BGM,Lubinsky,Stahl1998}, in particular
the convergence in capacity of Padé approximants implies that spurious
poles have a nearby zero when the order gets large, leading to so-called
near pole-zero cancellations (also known as Froissart doublets). Modifications
of Padé approximants were proposed to offset this issue \cite{GGT},
but they do not eliminate the problem \cite{BM}. Apparently, the
theoretically less studied vector fitting method which is a least
squares version of linear interpolation, popular today in system analysis,
is also prone to producing spurious poles and near cancellations (see
\cite{LA} for issues on convergence of this method). 

In system identification, near cancellations are often ascribed to
overmodelling. The terminology suggests an analogy with the stochastic
identification paradigm: though measurements may not correspond to
a rational transfer function, the basic assumption is that they arise
from a well-defined rational system R with added noise. This point
of view leads one to postulate the existence of a \textquotedblleft correct
order\textquotedblright{} to identify the \gls{FRF}, i.e. the degree
of R, while using a higher degree results in approximating the noise
term with inessential, nearly simplifying rational elements. However,
if the transfer function is not rational, requirements to keep the
degree small conflict with the need to make the approximation error
small as well (not to incur undermodelling), thus calling for a compromise
akin to the classical trade-off between bias and variance from parametric
stochastic identification \cite{Ljung}. To quote \cite{Ankabe2010}\footnote{The rational approximation technique used in this reference is described
as \textquotedblleft frequency domain least squares identification\textquotedblright{}}: \textquotedblleft it is not always trivial to discriminate between
overmodelling quasi-cancellations and physical quasi- cancellations
that really reflect an unstable behaviour\textquotedblright . 

To resolve this issue, the approach proposed in \cite{Ankabe2010}
is to cut the frequency band into smaller intervals and use low-order
local rational approximations to assess the stability of the \gls{FRF}
on each interval separately. On small enough intervals, rational approximation
can be performed accurately in low degree, and if unstable poles occur
their physical character is checked by re-modelling the \gls{FRF}
locally around each of them and verifying that the unstable pole remains
present in the new model. This procedure is commercially available
in the STAN tool \cite{STANtool,Mallet2009} and successful applications
on several examples are reported in \cite{vanHeijningen2016,Otegi2012,Collantes2011,Ayllon2011}.

Still, justifying the above-described technique presently rests on
heuristic arguments, and putting it to work is likely to require some
know-how since several parameters need to be adjusted adequately (Appendix
\ref{sec:LocalApproximation}, for example, shows that local models
of a stable \glspl{FRF} can become unstable). This is why the authors
feel that it may be interesting to develop an alternative viewpoint,
focusing more on estimating the unstable part of the gls{FRF}. 

Below, we propose a closed-loop stability analysis method devoid of
local models, in which the \gls{FRF} is projected onto the orthogonal
basis of stable and unstable functions. If a significant part of the
\gls{FRF} is projected onto the unstable basis functions, the circuit
solution is unstable. Calculating the projection boils down to computing
a Fourier transform once the \gls{FRF} is mapped from the imaginary
axis to the unit circle. Using the \gls{FFT}, this can be done fast
and in a numerically robust way. 

Functional projection onto a stable and unstable basis is a linear
operation, simple to implement, and no optimisation step is required.
No model-order or maximum approximation error needs to be specified.
The parameters in the projection method are the frequency range on
which the \gls{FRF} is determined and the amount of simulation points.
When the amount of simulated points is too low, an interpolation error
is present in the result of the method. It is shown that the level
of this error can easily be estimated and used to correctly choose
the amount of needed simulation points.

Once the unstable part of the \gls{FRF} has been obtained, it is
compared to the level of the interpolation error to determine whether
the unstable part is significant or not. This final step can be done
visually, or a significance threshold can be chosen by the user, both
will require some experience with the method. 

A final benefit of the projection-based approach is that it may help
exploiting the fact that the unstable part is rational in a small-signal
stability analysis \cite{Baratchart2014}. The unstable part can therefore
be approximated by a rational function without influence of the distributed
elements, which are projected onto the stable part of the \gls{FRF}.

The following of the paper is structured as follows: First, the simulation
set-up used to determine the \gls{FRF} of the linearised circuit
is discussed (Section~\ref{sec:FRF}). Then, the details of the functional
projection are provided (Section~\ref{sec:Stable/unstable-projection}).
In Section~\ref{sec:Examples}, the method is applied to four examples:
First, an artificial example is considered. Then, the small-signal
stability of two amplifiers is investigated and finally, the method
is applied to investigate the large-signal stability of a circuit.

\section{Determining the frequency responses\label{sec:FRF}}

\begin{figure}
\begin{centering}
\includegraphics{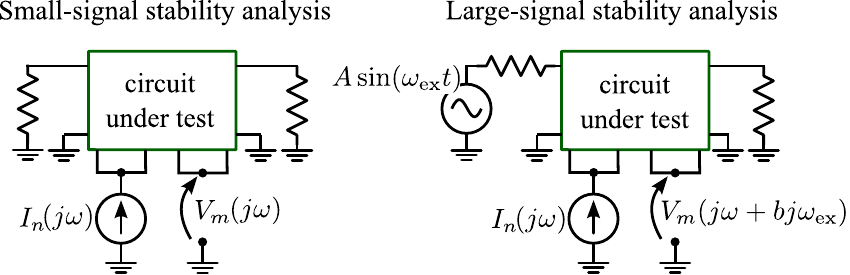} 
\par\end{centering}
\caption{\label{fig:simsetup}A Small-signal current source is connected to
a well-chosen node in the circuit under test to perform the local
stability analysis.}
\end{figure}
In this paper, the (trans)impedance presented by the circuit to a
small-signal current source will be used as \gls{FRF} (Fig.~\ref{fig:simsetup}).
In the remainder of this paper, it will be assumed that the unstable
poles are observable in the \gls{FRF}. To reduce the chance of
missing an instability in the circuit due to a pole-zero cancellation,
many different \glspl{FRF} can be analysed one-by-one. Having a
fast method to determine stability of a single \gls{FRF} is therefore
critical to a robust stability analysis.

The \gls{FRF} of the linearised circuit is obtained by first placing
the circuit in the required orbit, using either a DC or \gls{HB}
analysis and running a small-signal simulation around this orbit.

In a small-signal stability analysis, the stability of the DC solution
of the circuit is investigated, so the \gls{FRF} of the linearised
circuit is obtained with an AC simulation. The impedance of the circuit
is then obtained as: 
\begin{equation}
\imp=\frac{V_{m}\jw}{I_{n}\jw}\label{eq:impedance}
\end{equation}
where $I_{n}\jw$ is the small-signal current injected into the selected
node $n$ and $V_{m}\jw$ is the voltage response of the circuit measured
at node $m$ in the circuit.

In a large-signal stability analysis, the stability of a large-signal
solution of the circuit is investigated. The circuit is driven by
a periodic continuous-wave excitation at a pulsation $\omega_{\ex}$
and the circuit solution is obtained with a \gls{HB} simulation.

The \gls{FRF} of the linearised system around the \gls{HB} orbit
is obtained with a mixer-like simulation\footnote{In Keysight's \gls{ADS}, this mixer-like simulation is called a
\gls{LSSS} analysis.}. As the small-signal will mix with the large signal, several transfer
impedances with a different frequency translation are obtained: \global\long\def\imp#1{Z_{mn}^{\left[ #1 \right]}\jw}
\[
\imp b=\frac{V_{m}\!\left(j\omega+bj\omega_{\ex}\right)}{I_{n}\!\left(j\omega\right)}\quad b\in\mathbb{Z}
\]
The stability analysis now needs to determine whether the obtained
impedances have poles in the right half-plane. The stability analysis
of a large-signal orbit doesn't differ much from the analysis of a
DC solution \cite{Collantes2004}. The small-signal stability analysis
can be considered a special case where only $\imp 0$ is analysed.

\section{Stable/unstable projection\label{sec:Stable/unstable-projection}}

The projection described here has been used before to perform stable
interpolation and extrapolation of \gls{FRF} data~\cite{Olivi2013}.
With slight modifications, it can be turned into a full-blown stability
analysis. We first start with a brief introduction to the notion of
Hardy spaces.

The Hardy space $H^{2}(\mathbb{C}^{+})$, is defined as the set of
all functions $g$ defined on $\mathbb{C}^{+}$ such that: 
\begin{itemize}
\item $\forall z\in\mathbb{C}^{+}$, $g$ is holomorphic at $z$ 
\item $\sup_{x>0}\int_{-\infty}^{+\infty}|g(x+j\omega)|^{2}d\omega<\infty$ 
\end{itemize}
A classical result \cite{Hoffman,Rudin} states that every function
$g\in H^{2}$ admits a limiting function $G(j\omega)$ defined on
the imaginary axis $j\mathbb{R}$. The latter is obtained by taking
the limit of $g(z)$ when $z$ tends non tangentially toward $j\omega$.
Moreover $\forall z\in\mathbb{C}^{+}$, $g(z)$ is equal to the poisson
integral of $G$, that is: 
\begin{equation}
g(z=x+jy)=\int_{-\infty}^{+\infty}G(j\omega)\frac{x}{x^{2}+(y-\omega)^{2}}d\omega.\label{poisson}
\end{equation}
The holomorphic nature of $g$ ensures that it is also the Cauchy
integral of its boundary value $G$: 
\begin{equation}
g(z=x+jy)=\frac{1}{2\pi}\int_{-\infty}^{+\infty}G(j\omega)\frac{1}{j\omega-z}d\omega.\label{cauchy}
\end{equation}
There is a one to one linear correspondence between the Hardy functions
$g$ and its boundary value function $G$. Using this identification,
$H^{2}(\mathbb{C}^{+})$ becomes a subspace of the Hilbert space $L^{2}\!\left(j\mathbb{R}\right)$
of square integrable functions on $j\mathbb{R}$. A direct consequence
of (\ref{poisson}) is that $H^{2}(\mathbb{C}^{+})$ is closed in
$L^{2}\!\left(j\mathbb{R}\right)$ and therefore admits an orthogonal
complement. An important result \cite{Hoffman} asserts that, $\left(H^{2}(\mathbb{C}^{+})\right)^{\perp}=H^{2}(\mathbb{C}^{-})$
where $H^{2}(\mathbb{C}^{-})$ is defined exactly as $H^{2}(\mathbb{C}^{+})$
above by replacing $\mathbb{C}^{+}$ by $\mathbb{C}^{-}$ and taking
the supremum over $x<0$. We therefore have that 
\[
L^{2}\!\left(j\mathbb{R}\right)=H^{2}\!\left(\mathbb{C}^{+}\right)\oplus H^{2}\!\left(\mathbb{C}^{-}\right)
\]
This decomposition asserts that any square integrable function on
$j\mathbb{R}$ decomposes uniquely as the sum of the traces on the
imaginary axis, of an analytic function in the right half-plane and
a function analytic in the left half-plane. The projection on $H^{2}\!\left(\mathbb{C}^{+}\right)$
defines the stable part of the function. The projection onto $H^{2}\!\left(\mathbb{C}^{-}\right)$
is the unstable part. As an example, consider $P/Q$ a strictly proper
($\deg(P)<\deg(Q)$) rational function devoid of poles on the imaginary
axis. We write its partial fraction expansion as, 
\[
\frac{P(s)}{Q(s)}=\sum_{i\in I^{+}}\sum_{k=1}^{k_{i}}\frac{a_{i,k}}{(s-\lambda_{i})^{k}}+\sum_{i\in I^{-}}\sum_{k=1}^{k_{i}}\frac{a_{i,k}}{(s-\lambda_{i})^{k}}
\]
where the $\lambda_{i}'s$ with $i\in I^{-}$ are poles belonging
to $\mathbb{C}^{-}$, and the ones with $i\in I^{+}$ belong to $\mathbb{C}^{+}$.
The strict properness of $P/Q$ ensures its square integrability on
$j\mathbb{R}$. By unicity its stable part obtained after projection
on $H^{2}\!\left(\mathbb{C}^{+}\right)$ is found to be, 
\[
\sum_{i\in I^{-}}\sum_{k=1}^{k_{i}}\frac{a_{i,k}}{(s-\lambda_{i})^{k}},
\]
while its unstable part is 
\[
\sum_{i\in I^{+}}\sum_{k=1}^{k_{i}}\frac{a_{i,k}}{(s-\lambda_{i})^{k}}.
\]

In the general case the projection boils down to calculating certain
inner products of $\imp b$ with the basis functions $B_{k}$, which
form an orthogonal basis of $L^{2}\!\left(j\mathbb{R}\right)$ 
\begin{align}
c_{k}=\left\langle \imp b,B_{k}\right\rangle = & \intop_{-\infty}^{\infty}\imp b\overline{B_{k}\jw}d\omega\label{eq:scalarproduct}\\
B_{k}\s= & -\sqrt{\frac{\alpha}{\pi}}\frac{\left(s-\alpha\right)^{k}}{\left(s+\alpha\right)^{k+1}}\qquad k\in\mathbb{Z}\label{eq:splanebasis}
\end{align}
The overbar $\overline{\bullet}$ indicates the complex conjugate.
$\alpha$ is a positive constant used for scaling. All $B_{k}$ with
$k\geq0$ create a basis for the stable part, while the $B_{k}$ with
negative $k$ form a basis for $H^{2}\!\left(\mathbb{C}^{-}\right)$.
Once the $c_{k}$ coefficients are calculated, the stable and unstable
parts are easily recovered by calculating 
\begin{flalign}
\Zstable & =\sum_{k=0}^{\infty}c_{k}B_{k}\jw\label{eq:stablePart}\\
\Zunstable & =\sum_{k=1}^{\infty}c_{-k}B_{-k}\jw\label{eq:unstablePart}
\end{flalign}
The inner product in \eqref{eq:scalarproduct} runs over all frequencies
while the impedance function $\imp b$ is only known over a frequency
range $\mathbf{f}=\left[\fmin,\fmax\right]$. To impose the finite
frequency band on the data, the impedance is filtered before the analysis
\begin{equation}
\Zfiltered=\imp b\filter\label{eq:filtering}
\end{equation}
The filter $\filter$ is a high-order elliptic lowpass filter with
its first transmission zero placed at $\fmax$ that imposes band limitation.
This filter will stabilise poles close to its cutoff frequency, so
$\fmax$ should be chosen well beyond the maximum frequency at which
the circuit can become unstable. $\fmin$ should be placed very close
to DC. If $\fmin$ can't be close to DC, a bandpass filter should
be used for $\filter$. The filtering ensures that $\Zfiltered\in L^{2}\!\left(j\mathbb{R}\right)$
by suppressing anything outside of the frequency band of interest.
The smooth decay to zero of $\Zfiltered$ at the edges of the frequency
interval will avoid instabilities to pop up due to the discontinuity
of $\imp b$.

In this paper, we use an elliptic filter of order 10 to filter the
data. The filter has one transmission zero which is placed exactly
at $\fmax$.

\subsection{Transforming to the unit circle}

\begin{figure}
\begin{centering}
\includegraphics{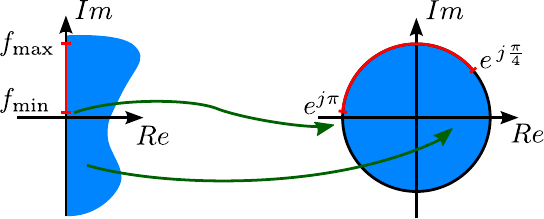} 
\par\end{centering}
\caption{\label{fig:MobiusTransform}The Möbius transform used in the analysis
maps the complex right-half plane to the inside of the unit disc.
DC is mapped to $-1$. When $\alpha=2\pi f_{\mathrm{max}}\left(\sqrt{2}-1\right)$,
$\protect\fmax$ is mapped to $e^{j\frac{\pi}{4}}$.}
\end{figure}

\noindent Working with the basis functions defined in \eqref{eq:splanebasis}
is troublesome from a numeric point of view. Performing the projection
when working on unit disc yields better results \cite{Olivi2013}.
The mapping from the complex plane to the unit circle is performed
with the Möbius mapping visualised in Fig.~\ref{fig:MobiusTransform}:
\begin{equation}
Z_{\mathrm{f}}\!\left(s\right)\underset{\mathrm{M\ddot{o}b}}{\longmapsto}Z_{\mathrm{f}}^{\mathrm{disc}}=\sqrt{\pi\alpha}\frac{2}{z-1}Z_{\mathrm{f}}\left(\alpha\frac{1+z}{1-z}\right)\label{eq:transfo}
\end{equation}
Our mapping of choice converts square integrable functions on the
frequency axis into square integrable functions of the same norm on
the unit circle. Square integrable functions which are analytic on
the right half-plane are mapped onto square integrable functions which
are analytic inside the unit disc. Appendix~\ref{App:BasisTransfo}
shows that the basis functions $B_{k}$ map onto powers of $z$ 
\begin{equation}
B_{k}\!\left(s\right)\underset{\mathrm{M\ddot{o}b}}{\longmapsto}B_{k}^{\mathrm{disc}}\!\left(z\right)=z^{k}
\end{equation}
Projecting on this basis boils down to calculating the Fourier series
of $Z_{\mathrm{f}}^{\mathrm{disc}}\!\left(z\right)$, with coefficients
given by 
\begin{equation}
c_{k}=\frac{1}{2\pi}\intop_{0}^{2\pi}Z_{\mathrm{f}}^{\mathrm{disc}}\left(e^{j\theta}\right)e^{-jk\theta}d\theta
\end{equation}
This Fourier series can be calculated in a numerically efficient way
using the \glsentryfirst{FFT}. The \gls{FFT} requires that the
$\theta$-values are linearly spaced between $0$ and $2\pi$. Due
to the mapping from the complex plane to the unit disc, the samples
will not satisfy this constraint. A simple interpolation, can be used
to obtain $Z_{\mathrm{f}}^{\mathrm{disc}}$ on the $\theta$-values
required to perform the \gls{FFT}.

The interpolation can introduce artefacts in the unstable part if
the \gls{FRF} is not sampled on a sufficiently dense frequency
grid, which will be shown on an example later. The level of this interpolation
error can be estimated by interpolating the \gls{FRF} using only
the data at the even data points. The difference between the original
and the interpolated data for the odd data points will give an indication
of the interpolation error encountered in the stability analysis.
When the unstable part lies significantly above the interpolation
error, we can conclude that the original impedance is unstable.

The threshold to determine when the unstable part lies significantly
above the level of the interpolation error will depend on the simulation
set-up for the circuit. In the examples that follow, a level of $20~\mathrm{dB}$
has been used as a threshold. A more strict threshold could be used
at the cost of requiring a more dense frequency grid and an increased
simulation time. When measured components are used in the simulation
set-up, the noise level in those measurements should be taken into
account to choose the correct threshold.

\subsection{Summary of the projection method}

The stable and unstable parts of a \gls{FRF} are determined using
the following steps: 
\begin{enumerate}
\item Multiply the \gls{FRF} with a high-order filter as in (\ref{eq:filtering}) 
\item Transform the filtered \gls{FRF} to the unit disc using (\ref{eq:transfo}) 
\item Interpolate the transformed \gls{FRF} to a linear grid and use
the \gls{FFT} to calculate the $c_{k}$ coefficients 
\item Reconstruct the stable and unstable part using (\ref{eq:stablePart}-\ref{eq:unstablePart}) 
\end{enumerate}

\subsection{Obtaining the unstable poles\label{sec:Fitting-poles}}

A function is meromorphic on $\mathbb{C}$ if it is holomorphic on
$\mathbb{C}$ but on a countable number of isolated poles. The function
$\tanh(\omega)$ is for example meromorphic, having infinitely many
isolated poles on the imaginary axis placed at $j\pi/2+jk\pi\,\,(\forall k\in\mathbb{Z})$
and being analytic elsewhere. In a small-signal stability analysis
of a circuit composed of lumped elements, transmission lines and active
devices modelled by negative resistors, the impedances can be shown
to be meromorphic functions of the frequency. Under the additional
realistic assumption that active elements can only deliver power over
a finite bandwidth, the impedances are proven to possess only finitely
many unstable poles in $\mathbb{C}^{+}$ \cite{Baratchart2014}. Under
the generic condition that $\imp b$ is devoid of poles on the imaginary
axis, and that the filtering function $\filter$ decays strongly enough
in order to render $\imp b\filter$ square integrable, we conclude
that the unstable part of $\imp b\filter$ is a rational function.
Its poles coincide with the unstable poles of $\imp b$: note here
that the multiplication by $\filter$ does not add any unstable pole
as $\filter$ is stable. This means that most of the complexity of
the frequency response, like the delay, will be projected onto the
stable part, while the unstable part can easily be approximated by
a low-order rational model to recover the unstable poles\footnote{Interpolation error will be present in the obtained unstable part,
but its influence can be minimised by weighting the rational estimator
with the obtained interpolation error level.}.

Classic rational approximation tools can be used to approximate the
unstable part and determine the unstable poles. When multiple frequency
responses are analysed simultaneously, an approximation method suited
for approximation of rational matrices is preferred~\cite{RARL2}.
In our current implementation, Kung's method \cite{Kung1978,Markovsky2012}
is used to estimate the poles of the unstable part of a single \gls{FRF}
at a time. Alternatively, more sophisticated rational approximation
engines like RARL2~\cite{RARL2} can be used to recover and track
unstable poles.

Compared to working with a high-order rational approximation of the
total impedance of the circuit, the \lq{split-first, approximate
later}\rq~approach proposed here could be a faster and easier method
to recover the unstable poles. The post-processing will be faster,
but the amount of points required to obtain a sufficiently low interpolation
error might be higher than the amount of points required for a tool
based on rational approximation. 

When the circuit is stable, designers often require information about
critical stable poles, to determine how far the circuit is from instability
and to track the location of the poles as the circuit varies. In its
current form, the projection-based analysis does not simplify finding
the location of the stable poles. To perform such an analysis, methods
based on local modelling may still be required.

\section{Examples\label{sec:Examples}}

The stability analysis will now be applied to four different examples.
The first is an artificial example generated in Matlab on which we
can demonstrate that the unstable poles in the circuit are recovered
perfectly. In the second example, an unstable balanced amplifier is
analysed to show that the method works for RF circuits. The third
example is a two-stage GaN \gls{PA}. In the final example, a large-signal
stability analysis is performed to verify the stability of a circuit
orbit obtained in a \gls{HB} simulation.

All simulations were performed in Keysight's \glsentryfirst{ADS}
and the post-processing was performed in Matlab.

\begin{figure}
\begin{centering}
\includegraphics{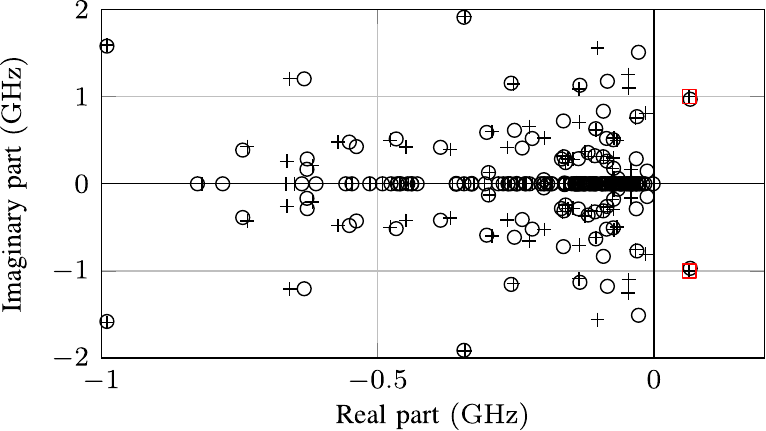} 
\par\end{centering}
\caption{\label{fig:Ex1_pzmap}Pole-zero map of the test system. There are
202 poles (\marker{black}{+}) and 200 zeroes (\marker{black}{o}).
The two poles placed in the right half-plane are easily recovered
after the projection by fitting a low-order model on the unstable
part (\marker{red}{square}).}

\begin{centering}
\includegraphics{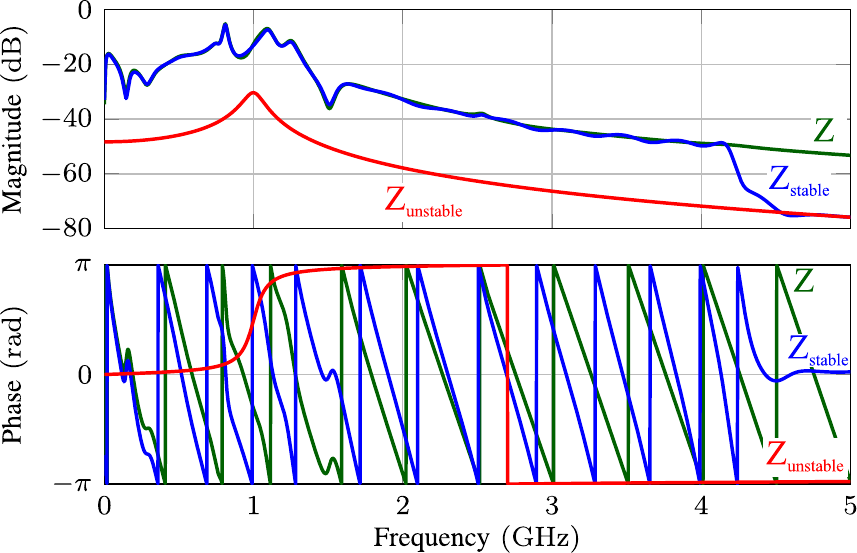} 
\par\end{centering}
\caption{\label{fig:Ex1_projres}The stable/unstable projection splits the
original frequency response (\marker{black!30!green}{-}) in a stable
part indicated with (\marker{blue}{-}) and an unstable part indicated
with (\marker{red}{-})}
\end{figure}

\subsection{Example 1: Random state space system}

As a first example, the stability analysis is applied to a random
system of order $202$ generated with the rss function from Matlab\footnote{The rss function in Matlab returns models with poles and zeroes around
$1\mathrm{Hz}$. The example here was scaled up in frequency to represent
an RF circuit.}. The test system has an unstable pole pair at $1~\mathrm{GHz}$,
as can be seen on its pole-zero map (Fig.~\ref{fig:Ex1_pzmap}).
A zero is placed close to the unstable poles. This makes that the
unstable poles are difficult to observe in the \gls{FRF}. To introduce
delay in the test system, a time delay of $2~\mathrm{ns}$ is added
to the system. The frequency response of the system is calculated
on 5000 linearly spaced frequency points between $0~\mathrm{Hz}$
and $5~\mathrm{GHz}$ and is shown in green in Fig.~\ref{fig:Ex1_projres}.
The obtained stable and unstable parts after projection are shown
in red and blue on the same figure. The maximum interpolation error
is very low in this example ($-120~\mathrm{dB}$) . We will focus
on the effect of the interpolation error in more detail in example
2.

The obtained unstable part peaks at $1~\mathrm{GHz}$, which matches
the location of the unstable pole pair of the system. Note also that
the obtained $Z_{\mathrm{unstable}}$ is very simple: it is clearly
a second-order system. Most of the complexity of the frequency response,
including the delay, is projected onto the stable part. This observation
supports the proposed approach of estimating a rational model only
after projection. A good fit was obtained with a rational model that
consisted of two unstable poles and a single zero. The two poles obtained
with a rational approximation of $Z_{\mathrm{unstable}}$ coincide
exactly with the unstable poles in the circuit as is shown in Fig.~\ref{fig:Ex1_pzmap}.

\begin{figure*}[h!]
\begin{centering}
\includegraphics[scale=0.9]{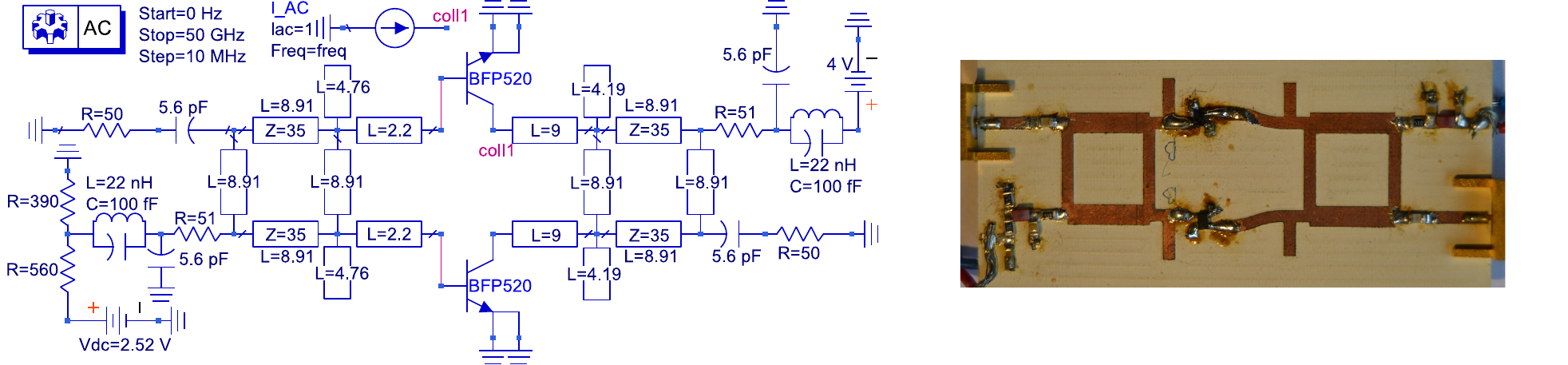}
\par\end{centering}
\caption{\label{fig:BALcircuit}Simulation set-up and photograph of the balanced
amplifier. In the simulations for the stability analysis, the amplifier
is excited at the collector of one of its transistors. All transmission
lines in the circuit are $50\Omega$ lines unless stated otherwise.
The length of the transmission lines is given in millimetres. The
TLINP model was used for the transmission lines with $\epsilon_{r}=6.15$,
$\tan(\delta)=0.003$ and conductor losses $A=2.5~\nicefrac{\mathrm{dB}}{\mathrm{m}}$.}

\begin{centering}
\includegraphics{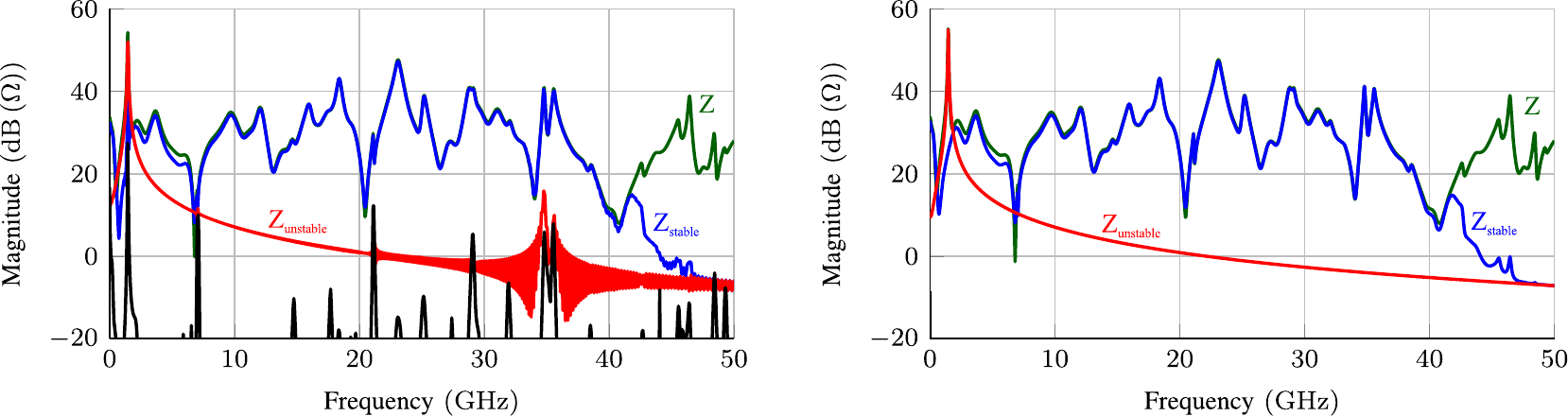} 
\par\end{centering}
\caption{\label{fig:BalAmpStabAnal}The separation of the impedance (\marker{black!30!green}{-})
into the stable part (\marker{blue}{-}) and unstable part (\marker{red}{-})
reveals the instability around $1.46~\mathrm{GHz}$. The interpolation
error is shown with (\marker{black}{-}). In the plot on the left,
there are artefacts present in the unstable part due to interpolation
of the coarsely obtained impedance data. When the impedance of the
balanced amplifier is simulated on a finer frequency grid, the artefacts
disappear (right).}
\end{figure*}

\subsection{Example 2: Balanced amplifier\label{sub:BAL}}

The second example is a balanced amplifier built as a student project
for operation around $3.4~\mathrm{GHz}$ (Fig.~\ref{fig:BALcircuit}).
Two BFP520 transistors were used to construct the amplifier. During
measurement, the design oscillated around $1.43~\mathrm{GHz}$ when
terminated with $50~\mathrm{\Omega}$, so the circuit is a good candidate
to verify the proposed method to find the instability in simulations.

The small-signal current source was connected to the collector of
the top transistor. The BFP520 has a $f_{\mathrm{T}}$ of $45~\mathrm{GHz}$,
so the maximum frequency for the simulation was set to $50~\mathrm{GHz}$.
The impedance of the circuit was determined starting from DC in $10~\mathrm{MHz}$
steps. The obtained impedance is shown in green in Fig.~\ref{fig:BalAmpStabAnal},
the obtained stable and unstable parts are also shown in the same
figure. The instability around $1.46~\mathrm{GHz}$ is detected, but
also some artefacts can be observed in the obtained unstable part
at higher frequencies. The high level of the interpolation error at
the frequency of these artefacts indicates that they are due to the
interpolation in step 3 of the stable/unstable projection. At the
frequency of the detected instability, the unstable part lies about
$30~\mathrm{dB}$ above the error level, which indicates that the
instability is not an interpolation artefact.

To confirm that the artefacts are caused by the interpolation, a second
simulation was run, but now $1~\mathrm{MHz}$ steps were used instead
of $10~\mathrm{MHz}$. The stable/unstable projection of the denser
frequency response data (Fig.~\ref{fig:BalAmpStabAnal}) still predicts
the instability around $1.46~\mathrm{GHz}$, but the artefacts in
the unstable part at higher frequencies are gone. The maximum of the
interpolation error went down to $-80~\mathrm{dB(\Omega)}$.

\subsection{Example 3: Two-stage Power Amplifier\label{sub:PA}}

\begin{figure}[tb]
\begin{centering}
\includegraphics[width=0.8\columnwidth]{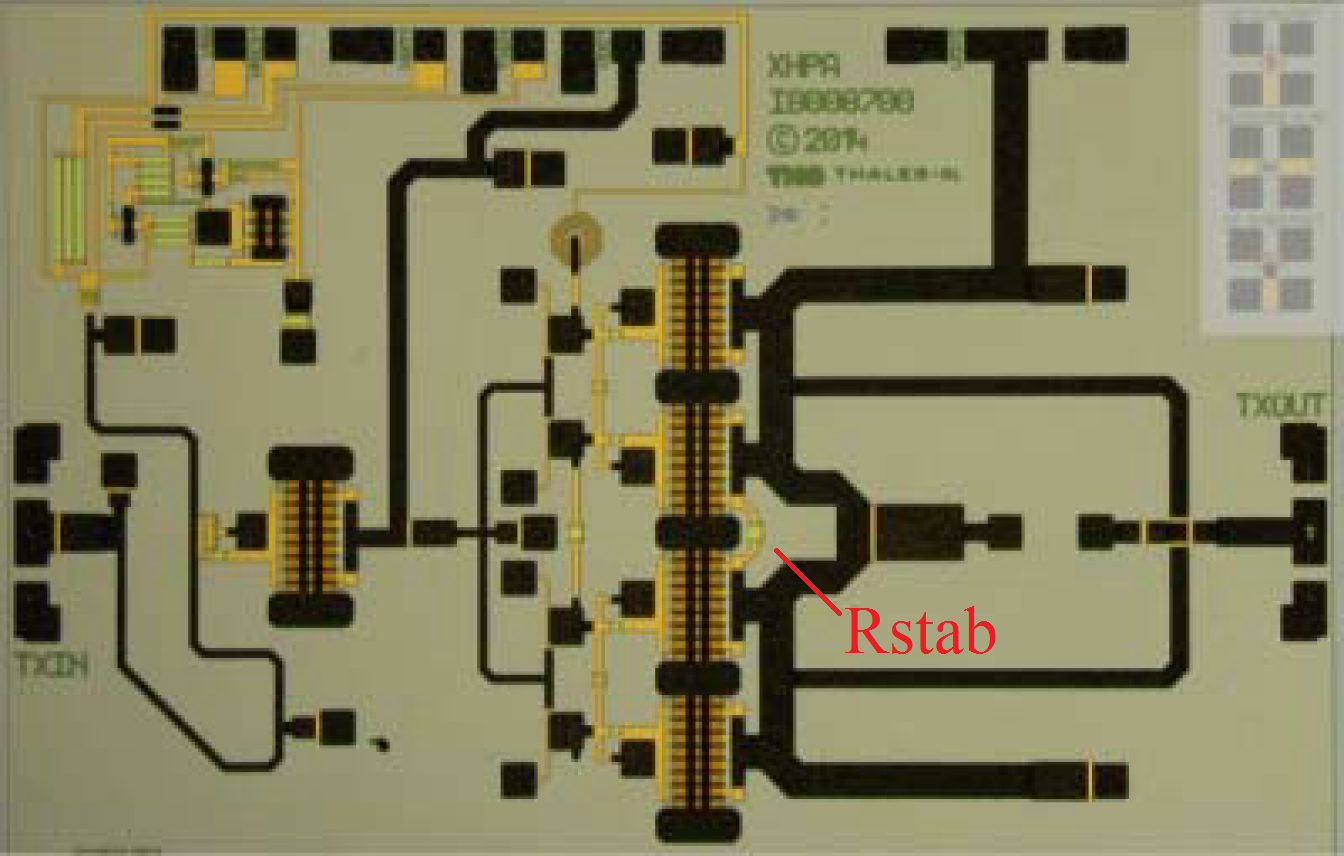}
\par\end{centering}
\caption{\label{fig:PAchip}Microphotograph of the MMIC. The stabilisation
resistor $\protect\Rstab$ is indicated in red.}

\vspace{1mm}

\begin{centering}
\includegraphics{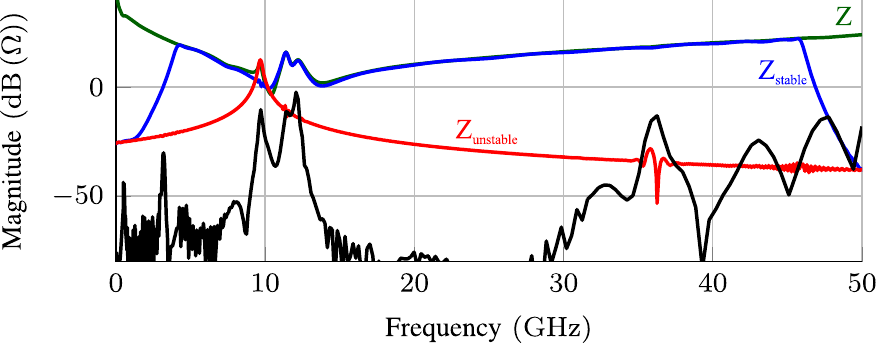} 
\par\end{centering}
\caption{\label{fig:PA_Rstab_500}Impedance seen at the gate of the first stage
of the \gls{PA} for $\protect\Rstab=500\Omega$. Its obtained stable
and unstable parts clearly indicate that the DC solution of the amplifier
is unstable. The interpolation error (\marker{black}{-}) is quite
high due to the low amount of simulation points.}

\begin{centering}
\includegraphics{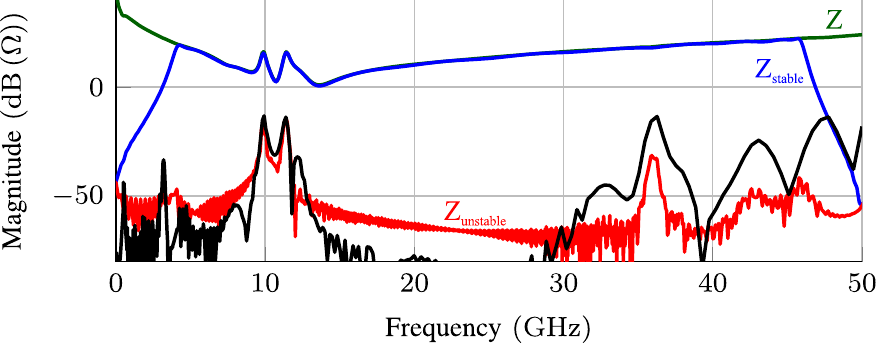} 
\par\end{centering}
\caption{\label{fig:PA_Rstab_25}Impedance presented by the \gls{PA} at
the gate of the transistor in the first stage for $\protect\Rstab=25\Omega$.
The obtained stable and unstable parts indicate that the DC solution
is now stable. The interpolation error is shown with (\marker{black}{-}).}
\end{figure}

As a third example, we consider the small-signal stability analysis
of an X-band \gls{PA} designed in the $0.25\mathrm{\mu m}$ GaN
HEMT technology GH25-10 of UMS~\cite{Floriot2012}. The circuit and
its design are described in great detail in \cite{vanHeijningen2016}.
The resulting MMIC is shown in Fig.~\ref{fig:PAchip}.

The \gls{PA} is a two-stage design where the second stage consists
of two branches with each two transistors in parallel. In simulation,
the second stage of the \gls{PA} demonstrated an odd-mode instability~\cite{vanHeijningen2016},
so a stabilisation resistor was added between the drains of the top
and bottom halves of the second stage of the \gls{PA} (as indicated
in the Fig.).

The simulation of the complete \gls{PA} was performed in \gls{ADS}.
The passive structures in the circuit were simulated with EM simulations
in Momentum and combined with the non-linear transistor models afterwards.
To verify the stability of the amplifier, the circuit impedance was
determined at the gate of the top most transistor of the second stage.

The obtained impedance for $\Rstab=500\Omega$ is shown in Fig.~\ref{fig:PA_Rstab_500}.
The impedance is simulated on 945 logarithmically spaced points between
$1\mathrm{MHz}$ and $50\mathrm{GHz}$. Because $1\mathrm{MHz}$ is
not sufficiently close to DC, a bandpass filter was used in the stability
analysis. Due to the low amount of data points in the resonances of
the circuit, a Padé interpolation was used in the stability analysis.
The resulting stable and unstable parts are shown in blue and red
on the same figure. It is clear that the circuit is unstable for $\Rstab=500\Omega$.
The unstable part peaks around $9.5\mathrm{GHz}$ and lies about $40~\mathrm{dB}$
above the interpolation error level.

The odd-mode instability can be resolved by decreasing the resistance
of $\Rstab$~\cite{vanHeijningen2016}. In a second stability analysis,
we determined the stability of the \gls{PA} for $\Rstab=25\Omega$.
The results of this second analysis are shown in Fig.~\ref{fig:PA_Rstab_25}.
The obtained unstable part coincides with the level of the interpolation
error, which indicates that the circuit is now stable.

\subsection{Example 4: R-L-diode circuit\label{sub:LSexample}}

\begin{figure*}[!htb]
\begin{minipage}[c]{0.55\textwidth}%
 \includegraphics{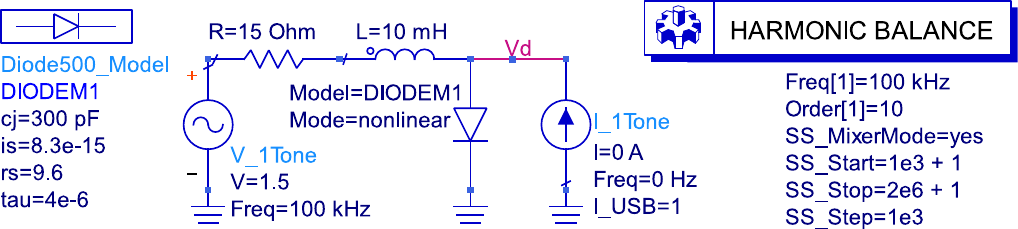}\caption{\label{fig:RLdiodecircuit}The circuit R-L-diode circuit is excited
with a small-signal current source at the diode.}
\end{minipage}\hfill{}%
\begin{minipage}[c]{0.4\textwidth}%
 \includegraphics[width=1\columnwidth]{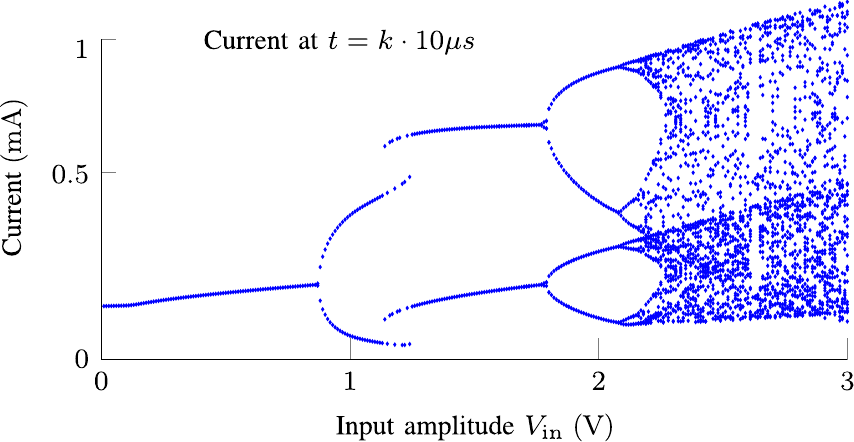}
\caption{\label{fig:bifurcationdiagram} The bifurcation diagram of the R-L-diode
circuit shows that a period-doubling occurs for input amplitudes higher
than $0.8~\mathrm{V}$.}
\end{minipage}
\begin{centering}
\includegraphics{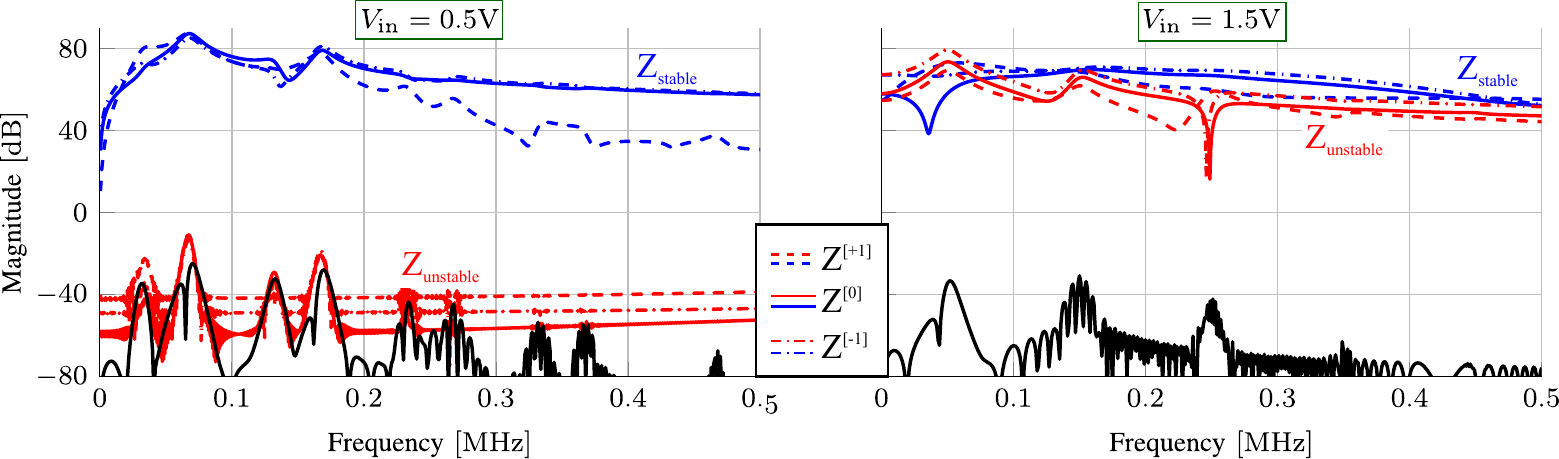} 
\par\end{centering}
\caption{\label{fig:LSresult}The results of the stability analysis of $\protect\imp{-1}$
$\protect\imp 0$ and $\protect\imp{+1}$ in the two \gls{HB} simulations
show that, for $V_{\mathrm{in}}=0.5~\mathrm{V}$, the orbit is stable,
but for $V_{\mathrm{in}}=1.5~\mathrm{V}$, the orbit is unstable.
The interpolation error in shown with (\marker{black}{-}).}
\end{figure*}

The final example in this paper shows that the stability analysis
can also be used to determine the stability of \gls{HB} simulations
of the R-L-diode circuit shown in Fig.~\ref{fig:RLdiodecircuit}.
The circuit is based on~\cite{Azzouz1983}, but a realistic diode
model was used to represent the diode in the circuit instead of the
three equations provided in the original paper.

The circuit is excited by a single-tone voltage source with an amplitude
$V_{\mathrm{in}}$ and a frequency of $100~\mathrm{kHz}$. Because
the diode has a transit-time of $4~\mathrm{\mu s}$, the circuit generates
period-doubling solutions starting from sufficiently high amplitudes
$V_{\mathrm{in}}$. For even higher $V_{\mathrm{in}}$, the circuit
will create chaotic solutions.

To visualise this behaviour, a bifurcation diagram is constructed
using time-domain simulations in the same way as is described in \cite{Azzouz1983}:
For every value of $V_{\mathrm{in}}$, 1030 periods of $100~\mathrm{kHz}$
are simulated and the final 30 periods are sampled every $\nicefrac{1}{100~\mathrm{kHz}}$.
If the circuit solution is periodic with the same period as the input
source, all 30 sampled points will fall on top of each-other. If a
period-doubling occurs in the circuit, two different values will be
obtained.

The obtained bifurcation diagram for our R-L-diode example is shown
in Fig.~\ref{fig:bifurcationdiagram}. It is clear that a period-doubling
occurs for $V_{\mathrm{in}}$ higher than $0.8~\mathrm{V}$. Starting
from $1.8~\mathrm{V}$ the period quadruples. For the highest input
amplitudes, a chaotic solution is obtained.

If this R-L-diode circuit is simulated with \gls{HB}, the circuit
solution is constrained to harmonics of $100~\mathrm{kHz}$. For input
amplitudes higher than $0.8~\mathrm{V}$, where the circuit wants
to go to a period-doubling solution, the constrained \gls{HB} solution
will be locally unstable.

We run two \gls{HB} simulations on this circuit. Both \gls{HB}
simulations have a base frequency of $100~\mathrm{kHz}$ and an order
of 10. In the first simulation, $V_{\mathrm{in}}$ is set to $0.5~\mathrm{V}$,
which will result in a stable orbit. The second simulation has a $V_{\mathrm{in}}$
of $1.5~\mathrm{V}$, which will cause the orbit to be unstable.

The frequency response of the circuit around the \gls{HB} solution
is obtained with a mixer-like simulation, as explained in the introduction
of this paper. The small-signal excitation was swept in both cases
on a linear frequency grid starting from $(1~\mathrm{kHz}+1~\mathrm{Hz})$
up to $(2~\mathrm{MHz}+1~\mathrm{Hz})$ in $1~\mathrm{kHz}$ steps.
The $1~\mathrm{Hz}$ was added to the start and stop values of the
sweep to avoid overlap with the tones of the \gls{HB} simulation.

The mixer-like simulation in \gls{ADS} uses \gls{SSB} current
excitations $i\left(t\right)=e^{j\omega t}$, which causes the obtained
frequency responses $Z_{mn}^{\prime\left[b\right]}\left(j\omega\right)$
with $b\neq0$ to be non-Hermitian: 
\[
Z_{mn}^{\prime\left[b\right]}\left(j\omega\right)\neq\overline{Z_{mn}^{\prime\left[b\right]}\left(-j\omega\right)}
\]
An alternative representation can make $Z_{mn}^{\prime\left[b\right]}\left(j\omega\right)$
Hermitian by transferring to a sine and cosine basis from the exponential
basis \cite{Louarroudi2014,Sandberg2005} 
\begin{flalign*}
\imp b & =\frac{1}{2}\left[Z_{mn}^{\prime\left[b\right]}\left(j\omega\right)+Z_{mn}^{\prime\left[-b\right]}\left(j\omega\right)\right]\\
\imp{-b} & =\frac{j}{2}\left[Z_{mn}^{\prime\left[b\right]}\left(j\omega\right)-Z_{mn}^{\prime\left[-b\right]}\left(j\omega\right)\right]
\end{flalign*}

$\imp{-1}$ $\imp 0$ and $\imp{+1}$ are then analysed with the stable/unstable
projection method. The results are shown in Fig.~\ref{fig:LSresult}.
The \gls{HB} solution obtained for $V_{\mathrm{in}}=0.5~\mathrm{V}$
is clearly stable: its unstable part is more than $70~\mathrm{dB}$
smaller than its stable part.

In the case for $V_{\mathrm{in}}=1.5~\mathrm{V}$, the solution is
clearly unstable as the unstable part lies far above the stable part
of the frequency response. Note that the lowest-frequency peak in
the unstable part is located around $50~\mathrm{kHz}$ and that copies
of the resonance are found at $150~\mathrm{kHz}$, $250~\mathrm{kHz}$,...
This behaviour is to be expected and indicates that the circuit wants
to go to a period-doubling solution.

During the stability analysis of a periodic orbit, the unstable part
will contain both the unstable base pole and all its higher-order
copies. The unstable part will be simple, just like in the small-signal
case, and it will be possible to approximate it by a finite set of
base poles. Due to the infinite amount of higher-order copies however,
it will not be possible to approximate it by a low-order rational
approximation as is the case in the stability analysis of a DC solution.

\section{Conclusion}

This paper introduces a closed-loop local stability analysis without
using a rational approximation. Instead, the impedance functions are
split into a stable and unstable part by projecting onto an orthogonal
basis. Transforming the problem to the unit disc allows to calculate
this projection with the \gls{FFT} which makes the projection-based
stability analysis very fast. In a small-signal stability analysis,
once the unstable part is obtained, a low-order rational model can
be used to find the unstable poles in the circuit.

Due to the model-free nature of the proposed method, it is a very
simple method to use: no choice of model order or approximation error
needs to be made. The only requirements of the projection-based stability
analysis are that the frequency responses are sampled on a sufficiently
dense frequency grid and that the maximum frequency of the simulations
is large enough. When the circuit impedance is simulated on a too
coarse frequency grid, a large interpolation error is introduced in
the results. The level of this interpolation error can easily be determined
and used to improve the accuracy of the method.

Once the stable and unstable parts of the impedance are obtained with
a sufficiently low interpolation error, the obtained unstable part
can be compared to the interpolation error to determine whether it
is significant or not. From experience, we found that, when the unstable
part lies more than $20~\mathrm{dB}$ above the interpolation error
level, the circuit can be considered unstable. Further work is to
be done towards automated decision making regarding stability.

The stable/unstable projection has been successfully applied to both
the stability analysis of DC and large-signal solutions of RF circuits.

\appendices{}

\section{On local rational approximation\label{sec:LocalApproximation}}

Due to the presence of distributed elements in microwave circuits
it is impossible to obtain a rational model of the impedances of the
circuit over the full simulated frequency range. In a small frequency
band however, it is possible to obtain a good low-order rational approximation
of the impedance. This is the basis of pole-zero based stability analysis. 

It is however not guaranteed that the poles of the local model correspond
to the poles of the global model. We will demonstrate this in this
appendix using an simple example without distributed elements.

\begin{figure*}[!]
\begin{equation}
Z\left(s\right)=\frac{\begin{array}{c}
-228.5s^{14}-153.7s^{13}-875.2s^{12}-550.2s^{11}-1364.9s^{10}-789.9s^{9}\\
-1118.6s^{8}-584.2s^{7}-520.9s^{6}-239.2s^{5}-140.7s^{4}-54.7s^{3}-21.4s^{2}-5.9s-1
\end{array}}{\nicefrac{\left(s+5\right)^{15}}{9.9281\cdot10^{10}}}\label{eq:AppFun}
\end{equation}

\centering{}\hrulefill
\end{figure*}

\begin{figure}
\setlength{\figurewidth}{\columnwidth}\setlength{\figureheight}{0.7\columnwidth}
\begin{centering}
\input{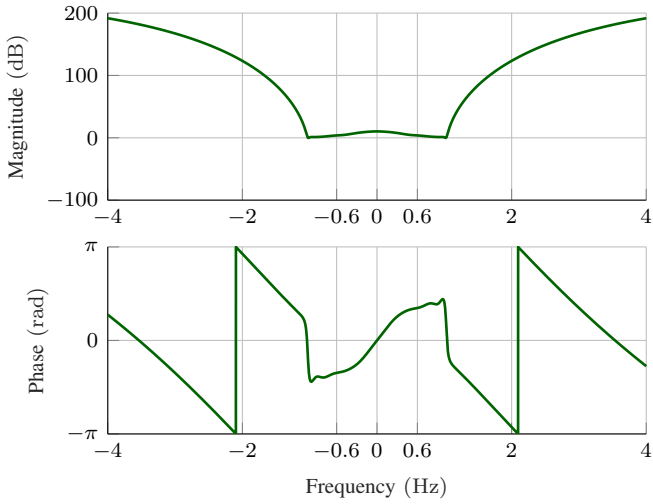}
\par\end{centering}
\caption{\label{App:FRF}\gls{FRF} used in Appendix A. The \gls{FRF} is stable,
but resembles an unstable \gls{FRF} in the interval {[}-1,1{]}. Using
a local model of this \gls{FRF} might result in false positives during
stability analysis.}
\end{figure}

Consider the artificial example in equation \eqref{eq:AppFun}. All
poles of this rational function lie in $s=-5$, so $Z(s)$ is stable.
In the frequency band $[-1,1]$ however, the \gls{FRF} of this stable
rational function closely resembles an unstable \gls{FRF} (Figure
\ref{App:FRF})

To verify whether the global poles are obtained with a local model,
we estimate a local rational model on 1000 frequency points on the
interval $\omega\in[-0.6,0.6]$ using vector fitting \cite{Gustavsen1999}.
The model order for the local model is not known in advance, so it
is swept from 2 to 11. The maximum of the phase error of the obtained
fit is shown in Figure \ref{App:fit}. Starting from model order 8,
the phase error lies below $10^{-3}$ degrees so the fit can be considered
sufficiently good for stability analysis \cite{Mallet2009}.

The obtained local model is unstable however. In fact, an unstable
local model is obtained for all model, which shows that the poles
of a local model do not necessarily correspond to the poles of the
underlying impedance. This is an artificial example of course, but
the example indicates that the use of local lower-order models to
determine the stability could lead to misleading results.

\begin{figure}
\setlength{\figurewidth}{0.8\columnwidth}\setlength{\figureheight}{0.4\columnwidth}
\begin{centering}
%
%
\definecolor{mycolor1}{rgb}{0.78431,0.78431,0.78431}%
\begin{tikzpicture}

\begin{axis}[%
width=\figurewidth,
height=\figureheight,
at={(0\figurewidth,0\figureheight)},
scale only axis,
xmin=2,
xmax=11,
xlabel style={font=\footnotesize\color{white!15!black}},
xlabel={\footnotesize model order},
ymin=-8,
ymax=-1,
ytick={-9,-6,-3,0},
yticklabels={{$10^{-9}$},{$10^{-6}$},{$10^{-3}$},{$10^{0}$}},
ylabel style={font=\footnotesize\color{white!15!black}},
ylabel={\footnotesize Phase error $\left[ \si{\degree} \right]$},
axis background/.style={fill=white},
xmajorgrids,
ymajorgrids
]
\addplot [color=mycolor1, line width=2.0pt, forget plot]
  table[row sep=crcr]{%
1	-1.51244545335199\\
2	-1.43325587084877\\
3	-1.95266508254331\\
4	-1.88002267052614\\
5	-2.14921307463317\\
6	-2.22087289685249\\
7	-2.80330578227869\\
8	-3.46094596905697\\
9	-5.5796964351128\\
10	-5.7769960979007\\
11	-7.94124444000273\\
};
\addplot [color=red, line width=2.0pt, only marks, mark=+, mark options={solid, red}, forget plot]
  table[row sep=crcr]{%
1	-1.51244545335199\\
2	-1.43325587084877\\
3	-1.95266508254331\\
4	-1.88002267052614\\
5	-2.14921307463317\\
6	-2.22087289685249\\
7	-2.80330578227869\\
8	-3.46094596905697\\
9	-5.5796964351128\\
10	-5.7769960979007\\
11	-7.94124444000273\\
};
\addplot [color=black, line width=1.0pt, forget plot]
  table[row sep=crcr]{%
1	-3\\
2	-3\\
3	-3\\
4	-3\\
5	-3\\
6	-3\\
7	-3\\
8	-3\\
9	-3\\
10	-3\\
11	-3\\
};
\end{axis}
\end{tikzpicture}%
\par\end{centering}
\caption{\label{App:fit}Maximum phase error as a function of the model order
obtained when estimating a local model in the range $\omega\in[-0.6,0.6]$
of the stable impedance \eqref{eq:AppFun}.}

\setlength{\figurewidth}{0.4\columnwidth}\setlength{\figureheight}{0.4\columnwidth}
\begin{centering}
\definecolor{mycolor1}{rgb}{0.78431,0.78431,0.78431}%
\begin{tikzpicture}

\begin{axis}[%
width=\figurewidth,
height=\figureheight,
at={(0\figurewidth,0\figureheight)},
scale only axis,
xmin=-0.8,
xmax=0.8,
xlabel style={font=\footnotesize\color{white!15!black}},
xlabel={\footnotesize real part $\left( \mathrm{rad} \right)$},
ymin=-0.8,
ymax=0.8,
ytick={  -1, -0.6,    0,  0.6,    1},
ylabel style={font=\footnotesize\color{white!15!black}},
ylabel={\footnotesize imaginary part $\left( \mathrm{rad} \right)$},
axis background/.style={fill=white},
axis x line*=bottom,
axis y line*=left,
xmajorgrids,
ymajorgrids,
legend style={legend cell align=left, align=left, draw=white!15!black}
]
\addplot [color=black, only marks, mark=x, mark options={solid, black}]
  table[row sep=crcr]{%
-0.187896873647519	0.157673595979969\\
-0.187896873647519	-0.157673595979969\\
0.42072266526479	0.155916058712043\\
0.42072266526479	-0.155916058712043\\
-0.161417912512232	0.456141636359485\\
-0.161417912512232	-0.456141636359485\\
0.278341284843475	0.496836483745743\\
0.278341284843475	-0.496836483745743\\
};
\addlegendentry{\footnotesize poles}

\addplot [color=black, only marks, mark=o, mark options={solid, black}]
  table[row sep=crcr]{%
0.694263312295042	4.44089209850063e-16\\
0.225271045188528	0.491392923138622\\
-0.152376623045868	0.456613005418241\\
-0.178260574470314	0.157571047040283\\
0.225271045188529	-0.491392923138624\\
-0.152376623045868	-0.456613005418241\\
-0.178260574470314	-0.157571047040283\\
};
\addlegendentry{\footnotesize zeros}

\addplot [color=red]
  table[row sep=crcr]{%
0.6	0\\
0.598792005883131	0.0380543517939387\\
0.595172887698477	0.0759554721442496\\
0.589157218357624	0.113550746616246\\
0.580769220837814	0.150688792308648\\
0.570042670644567	0.187220067419092\\
0.557020759809644	0.222997473396196\\
0.541755922971973	0.257876947253503\\
0.524309626241871	0.291718041660281\\
0.504752119698709	0.324384490473359\\
0.483162154518635	0.355744757432784\\
0.459626665871387	0.385672565811924\\
0.434240422863042	0.414047406889267\\
0.407105646934279	0.44075502519452\\
0.378331600250714	0.465687878575054\\
0.348034145742719	0.488745571230201\\
0.316335280566301	0.509835257969709\\
0.28336264486361	0.528872018068549\\
0.249249007801132	0.545779197212711\\
0.214131732955123	0.560488716159064\\
0.178152225196965	0.572941344866444\\
0.141455361305656	0.583086940994125\\
0.104188906600158	0.590884651807325\\
0.0665029199406067	0.596303078676752\\
0.0285491494942454	0.599320403509805\\
-0.00951957830088482	0.599924476604325\\
-0.0475499741140731	0.598112865571165\\
-0.085388902963971	0.59389286512856\\
-0.122884000839114	0.587281467728867\\
-0.159884288214021	0.578305295135965\\
-0.196240777990453	0.567000491228801\\
-0.231807075415877	0.553412576462749\\
-0.266439967563464	0.537596264574801\\
-0.3	0.519615242270663\\
-0.332352038319666	0.499541912780863\\
-0.3633658122826	0.477457104318499\\
-0.392916440367171	0.453449744612555\\
-0.420884932623793	0.427616502827318\\
-0.447158669805453	0.400061400309775\\
-0.471631856845672	0.370895391732363\\
-0.4942059488579	0.340235918317662\\
-0.514790047940986	0.308206434944044\\
-0.533301269192954	0.274935913036446\\
-0.549665074459242	0.240558321243968\\
-0.563815572471545	0.205212085995402\\
-0.575695784168698	0.169039534104858\\
-0.585257872131244	0.132186319671924\\
-0.592463333205837	0.0948008375840099\\
-0.597283153543851	0.0570336259825097\\
-0.599697925429911	0.0190367600988409\\
-0.599697925429911	-0.0190367600988407\\
-0.597283153543851	-0.0570336259825096\\
-0.592463333205837	-0.09480083758401\\
-0.585257872131244	-0.132186319671924\\
-0.575695784168698	-0.169039534104858\\
-0.563815572471545	-0.205212085995401\\
-0.549665074459242	-0.240558321243968\\
-0.533301269192954	-0.274935913036446\\
-0.514790047940986	-0.308206434944044\\
-0.4942059488579	-0.340235918317662\\
-0.471631856845673	-0.370895391732363\\
-0.447158669805453	-0.400061400309775\\
-0.420884932623793	-0.427616502827318\\
-0.392916440367171	-0.453449744612555\\
-0.3633658122826	-0.477457104318499\\
-0.332352038319666	-0.499541912780863\\
-0.3	-0.519615242270663\\
-0.266439967563464	-0.537596264574802\\
-0.231807075415877	-0.553412576462749\\
-0.196240777990453	-0.567000491228801\\
-0.159884288214021	-0.578305295135965\\
-0.122884000839114	-0.587281467728867\\
-0.0853889029639711	-0.59389286512856\\
-0.0475499741140733	-0.598112865571165\\
-0.00951957830088456	-0.599924476604325\\
0.0285491494942454	-0.599320403509805\\
0.0665029199406065	-0.596303078676752\\
0.104188906600158	-0.590884651807325\\
0.141455361305656	-0.583086940994125\\
0.178152225196965	-0.572941344866444\\
0.214131732955123	-0.560488716159064\\
0.249249007801132	-0.545779197212711\\
0.283362644863609	-0.528872018068549\\
0.316335280566301	-0.509835257969709\\
0.348034145742719	-0.488745571230201\\
0.378331600250713	-0.465687878575054\\
0.407105646934279	-0.44075502519452\\
0.434240422863042	-0.414047406889268\\
0.459626665871386	-0.385672565811924\\
0.483162154518635	-0.355744757432784\\
0.504752119698709	-0.324384490473358\\
0.524309626241871	-0.291718041660281\\
0.541755922971973	-0.257876947253503\\
0.557020759809643	-0.222997473396197\\
0.570042670644567	-0.187220067419092\\
0.580769220837814	-0.150688792308648\\
0.589157218357624	-0.113550746616246\\
0.595172887698477	-0.07595547214425\\
0.598792005883131	-0.0380543517939392\\
0.6	-1.46957615897682e-16\\
};

\addplot [color=black, forget plot]
  table[row sep=crcr]{%
0	-1\\
0	1\\
};
\addplot [color=black, forget plot]
  table[row sep=crcr]{%
-1	0\\
3	0\\
};
\end{axis}
\end{tikzpicture}%
\par\end{centering}
\caption{The obtained poles and zeroes for model order 8. The model was estimated
using data in the interval $\omega\in[-0.6,0.6]$. The red circle
indicates all points $s$ in the complex plane for which $\left|s\right|<0.6$.}
\end{figure}

\section{Mapping of the basis functions onto the unit disc}

\label{App:BasisTransfo}

Applying transform (\ref{eq:transfo}) to the basis functions of the
complex plane (\ref{eq:splanebasis}) yields the following: 
\begin{multline*}
B_{k}^{\mathrm{disc}}\left(z\right)=\sqrt{\pi\alpha}\frac{2}{z-1}B_{k}\left(\alpha\frac{1+z}{1-z}\right)\\
=-\sqrt{\pi\alpha}\frac{2}{z-1}\sqrt{\frac{\alpha}{\pi}}\frac{\left(\alpha\frac{1+z}{1-z}-\alpha\right)^{k}}{\left(\alpha\frac{1+z}{1-z}+\alpha\right)^{k+1}}=z^{k}
\end{multline*}

\newpage{}

\section*{Acknowledgement}

{\footnotesize{}This research was partly supported by the French space agency CNES, partly by the Flemish Agency for Innovation by Science and Technology (IWT-Vlaanderen) and partly by the Strategic Research Program of the VUB (SRP-19). 
We are also thankful to Juan-Marie Collantes (UPV) for fruitful discussions on the topic of closed loop stability analysis. 
We would like to thank Kurt Homan, Johan Nguyen and Dries Peumans for the design and measurement of the balanced amplifier.
Finally, we would like to thank Marc van Heijningen for providing the data of the MMIC PA. }{\footnotesize \par}

\bibliographystyle{IEEEtran}
\bibliography{PaperDatabase}

\begin{IEEEbiography}[{\includegraphics[height=1.25in]{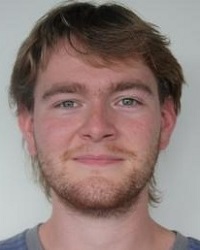}}]{Adam Cooman}
 was born in Belgium in 1989. He graduated as an Electrical Engineer
in Electronics and Information Processing in 2012 at Vrije Universiteit
Brussel (VUB) and obtained his Ph.D at the Department ELEC of the
VUB in December 2016. Now, Adam is part of the APICS team at INRIA,
Sophia Antipolis, France. His main interests are the design of Electronic
circuits, from low frequencies up to the microwave frequencies.
\end{IEEEbiography}

\begin{IEEEbiography}[{\includegraphics[width=1in]{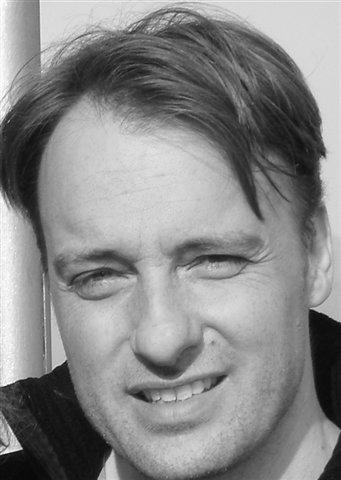}}]{Fabien Seyfert}
 graduated from the ``Ecole superieure des Mines'' (Engineering
School) in St Etienne (France) in 1993 and received his Ph.D in mathematics
in 1998. From 1998 to 2001 he was with Siemens (Munich, Germany) as
a researcher specialized in discrete and continuous optimization methods.
Since 2002 he occupies a full research position at INRIA (French agency
for computer science and control, Nice, France). His research interest
focuses on the conception of effective mathematical procedures and
associated software for problems from signal processing including
computer aided techniques for the design and tuning of microwave devices.
\end{IEEEbiography}

\begin{IEEEbiography}[{\foreignlanguage{english}{\includegraphics[width=1in]{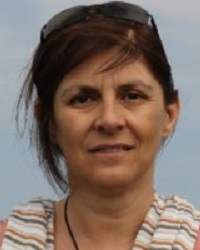}}}]{Martine Olivi}
 was born in France, in 1958. She got the engineer's degree from
Ecole des Mines de St-Etienne, France, and the PhD degree in Mathematics
from Université de Provence, Marseille, France, in 1983 and 1987 respectively.
Since 1988, she is with the Institut National de Recherche en Informatique
et Automatique (INRIA), Sophia Antipolis, France. Her research interests
include: rational approximation, parametrization of linear multivariable
systems, Schur analysis, identification and design of resonant systems.
Detailed information and publications are available at www-sop.inria.fr/members/Martine.Olivi
\end{IEEEbiography}

\begin{IEEEbiography}[{\includegraphics[width=1in]{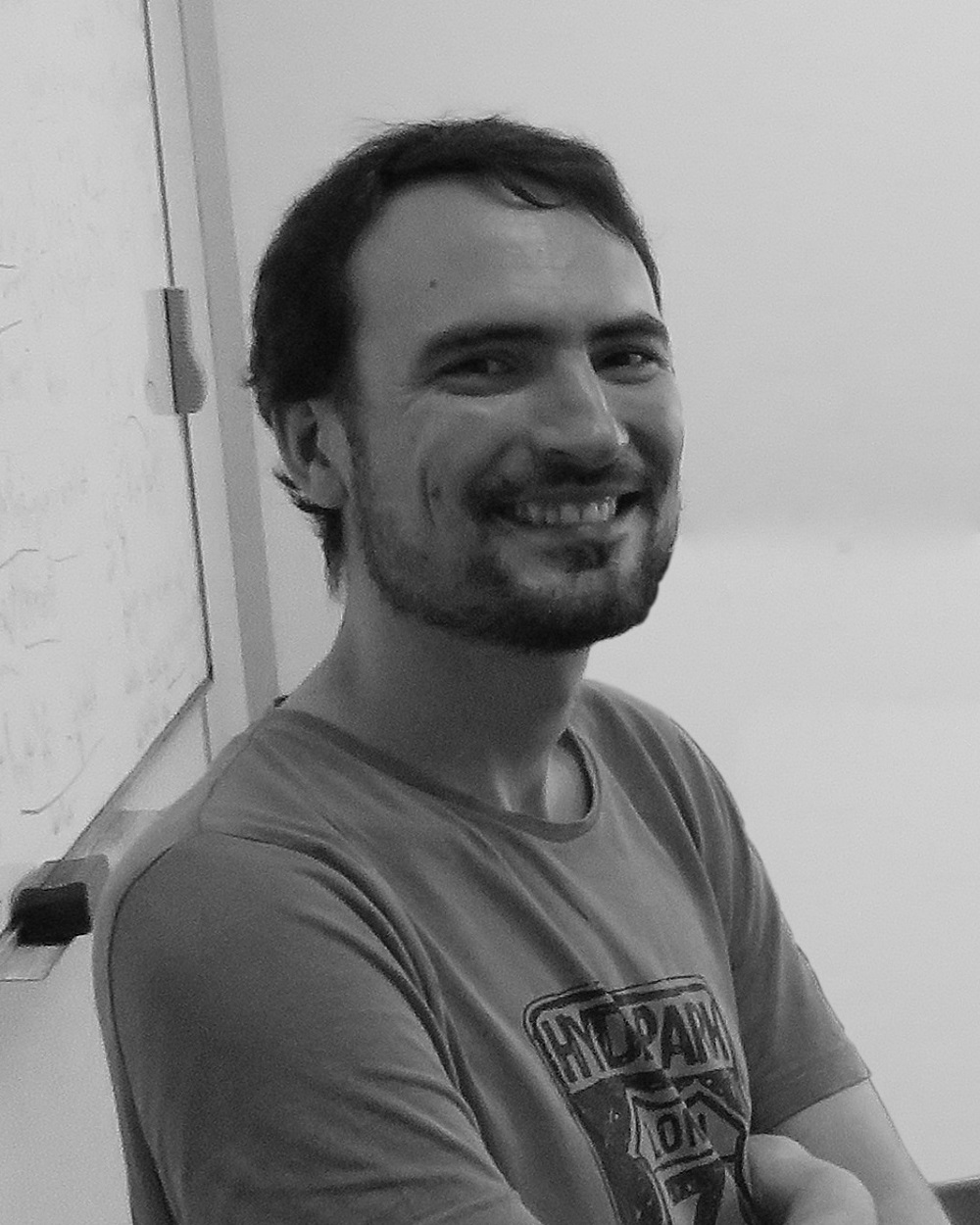}}]{Sylvain Chevillard}
 was born in Paris, France, in 1983. He received his Ph.D. in computer
science from Université de Lyon - École Normale Supérieure de Lyon
in 2009. He is Chargé de recherche (junior researcher) at Inria Sophia
Antipolis, France. Some of his research interests are reliable computing,
approximation theory, computer algebra, inverse problems.
\end{IEEEbiography}

\begin{IEEEbiography}[{\includegraphics[width=1in]{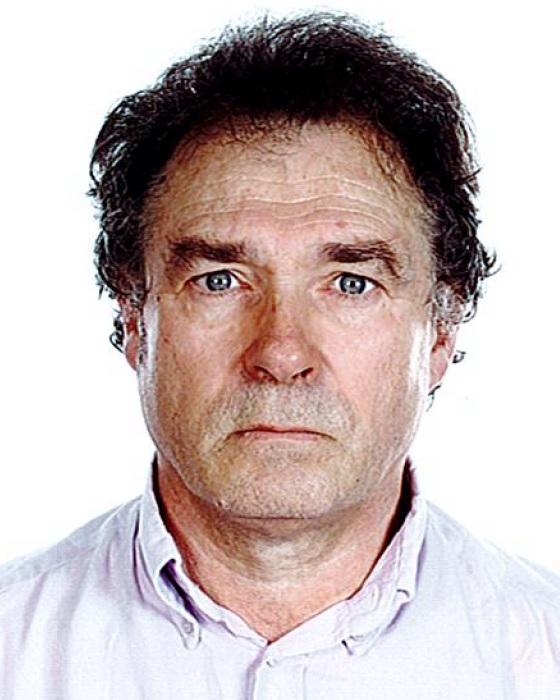}}]{Laurent Baratchart}
 received his Docteur Ingénieur degree from Ecole des Mines de Paris
in 1982 (advisor: Y. Rouchaleau) and his Thèse d'état in Mathematics
from the University of Nice in 1987 (advisor: A. Galligo). He was
the head of INRIA's project team MIAOU (Mathematics and Informatics
in Automatic control and Optimization for the User) from 1988 to 2003
and is currently the head of the project team APICS (Analysis of Problems
of Inverse type in Control and Signal processing) at INRIA Sophia-Antipolis
since 2004. His main interests lie with Complex and Harmonic Analysis,
Inverse Problems, as well as System and Circuit Theory.
\end{IEEEbiography}

\end{document}